\documentclass{article} 
\usepackage[final]{graphics}
\usepackage{amsfonts,amsbsy}

\font\tenmsa=msam10
\font\sevenmsa=msam7
\font\fivemsa=msam5
\font\tenmsb=msbm10
\font\sevenmsb=msbm7
\font\fivemsb=msbm5
\newfam\msafam
\newfam\msbfam
\textfont\msafam=\tenmsa  \scriptfont\msafam=\sevenmsa
\scriptscriptfont\msafam=\fivemsa
\textfont\msbfam=\tenmsb  \scriptfont\msbfam=\sevenmsb
\scriptscriptfont\msbfam=\fivemsb

\global\mathchardef\lesssim "142E

\newcommand{\slL}{\raise.15ex\hbox{$/$}\kern-.53em\hbox{$L$}}
\newcommand{\slP}{\raise.15ex\hbox{$/$}\kern-.53em\hbox{$P$}}
\newcommand{\slp}{\raise.1ex\hbox{$/$}\kern-.63em\hbox{$p$}}
\newcommand{\slq}{\raise.1ex\hbox{$/$}\kern-.63em\hbox{$q$}}
\newcommand{\slv}{\raise.1ex\hbox{$/$}\kern-.63em\hbox{$v$}}
\newcommand{\slR}{\raise.15ex\hbox{$/$}\kern-.53em\hbox{$R$}}
\newcommand{\slQ}{\raise.15ex\hbox{$/$}\kern-.53em\hbox{$Q$}}
\newcommand{\slK}{\raise.15ex\hbox{$/$}\kern-.53em\hbox{$K$}}
\newcommand{\slk}{\raise.15ex\hbox{$/$}\kern-.53em\hbox{$k$}}
\newcommand{\slSigma}{\raise.15ex\hbox{$/$}\kern-.53em\hbox{$\Sigma$}}
\newcommand{\slcalP}{\raise.15ex\hbox{$/$}\kern-.63em\hbox{$\cal P$}}
\newcommand{\slA}{\raise.15ex\hbox{$/$}\kern-.73em\hbox{$A$}}
\newcommand{\slbfA}{\raise.15ex\hbox{$/$}\kern-.73em\hbox{${\imb A}$}}
\newcommand{\slpartial}{\raise.15ex\hbox{$/$}\kern-.53em\hbox{$\partial$}}

\newcommand{\be}{\begin{equation}}
\newcommand{\ee}{\end{equation}}     
\newcommand{\bea}{\begin{eqnarray}}
\newcommand{\ena}{\end{eqnarray}}

\def\build#1\over#2{\mathrel{\mathop{\kern 0pt#1}\limits_{#2}}}

\font\tenimbf=cmmib10 at 10pt
\font\sevenimbf=cmmib10 at 7pt
\font\fiveimbf=cmmib10 at 5pt
\newfam\imbf
\textfont\imbf=\tenimbf
\scriptfont\imbf=\sevenimbf
\scriptscriptfont\imbf=\fiveimbf
\def\imb{\fam\imbf\tenimbf}

\begin{document}
\date{July 2001}
\title{\bf{Probing colored glass\\ via $q\bar{q}$ photoproduction}}
\author{F.~Gelis, A.~Peshier\\
Brookhaven National Laboratory,\\
Physics Department, Nuclear Theory,\\
Upton, NY-11973, USA}
\maketitle

\begin{abstract}
In this paper, we calculate the cross-section for the photoproduction
of quark-antiquark pairs in the peripheral collision of
ultra-relativistic nuclei, by treating the color field of the nuclei
within the Color Glass Condensate model. We find that this
cross-section is sensitive to the saturation scale $Q_s^2$ that
characterizes the model. In particular, the transverse momentum
spectrum of the produced pairs could be used to measure the properties
of the color glass condensate.
\end{abstract} 
\vskip 4mm 
\centerline{\hfill BNL-NT-01/15}

\section{Introduction}
In recent years, there has been a lot of interest in the description
of the distribution of partons inside a hadron or nucleus (see
\cite{Muell4,larry} for a pedagogical introduction). An outstanding issue is
the problem of the saturation of the parton distributions at very
small values of the momentum fraction $x$
\cite{GriboLR1,MuellQ1,FrankS1}. Indeed, the BFKL
\cite{Lipat1,KuraeLF1,BalitL1} equation predicts an uninterrupted rise
of this distribution as $x$ becomes smaller and smaller. However, this
is not compatible with unitarity, but rather an artifact of the linear
nature of this equation: in other words, the BFKL equation becomes
inadequate when the phase-space density approaches $1/\alpha_s$
\cite{GriboLR1,JalilKMW1,KovchM2}. This is an effect of the
nonlinearity of QCD, and happens even at small coupling.

Many of the recent investigations are based on a model introduced by
McLerran and Venugopalan \cite{McLerV1,McLerV2,McLerV3}, which
describes the small $x$ gluons inside a fast moving nucleus by a
classical color field. This approximation is justified by the large
occupation number of the soft gluon modes. In this model, the
classical Yang-Mills equation satisfied by the classical color field
is driven by the current induced by the hard partons. The distribution
of hard color sources inside the nucleus is described by a functional
density which is taken to be Gaussian in the simplest forms of this
model. This physical picture describes what can be called a ``Color
Glass Condensate''.  The classical version of this model depends only
on one parameter, called ``saturation scale'' and denoted $Q_s$,
defined as the transverse momentum scale below which saturation
effects start being important. This scale increases with energy and
with the size of the nucleus
\cite{Muell4,JalilKMW1,KovchM1}. Therefore, at energies high enough so
that $Q_s \gg \Lambda_{_{QCD}}$, the coupling constant at the
saturation scale is a small parameter and a perturbative approach can
be considered. In that case, ``perturbative'' means that one can work
at lowest order in $\alpha_s$, but all orders in the large classical
color field must be kept in principle.

Many theoretical improvements have dealt with the quantum corrections
to this model. When going to smaller values of $x$, one must integrate
out the modes comprised between the new scale and the former scale
\cite{JalilKMW1}.  Therefore, the functional density describing the
distribution of hard sources has been found to obey a functional
renormalization group equation \cite{JalilKLW1,JalilKLW2} that pilots
its variations when one moves the separation scale between the hard
and soft scale. Developments along those lines can be found in
\cite{JalilKW1,JalilKLW3,JalilKLW4,KovneM1,KovneMW3,Balit1,Kovch3}. A
particularly simple and suggestive form of this equation has been
found recently \cite{IancuLM1,IancuLM2}, as well as
approximate solutions \cite{IancuM1}.

On the phenomenological side, this model has been used in order to
study the fermionic degrees of freedom at small $x$ and compute
structure functions like $F_2$ \cite{McLerV4}. It has also been used
to study the production of gluons in the collision of two color
sources by several groups of
authors. \cite{KovneMW1,KovneMW2} calculated the production of gluons
at lowest order in the density of soft gluons. Then, Krasnitz and
Venugopalan \cite{KrasnV1,KrasnV2} attacked the problem of the gluon
production to all orders, by solving the classical Yang-Mills equation
on a 2-dimensional lattice for an SU(2) gauge theory. Recently, this
problem has been solved analytically for the case of $pA$ collisions
\cite{DumitM1}, for which the gluon density for one source is much
smaller than that of the second source. However, all these studies
deal with the central collision of two objects (AA, pA), and the
computed number of gluons cannot be observed directly. Instead,
several layers of complex processes are to be taken into account
between the stage where gluons are initially produced, and the stage
where hadrons are observed in a detector (in particular, the
hydrodynamical evolution, and the freeze-out of the system).

\begin{figure}[ht]
\centerline{\resizebox*{!}{3cm}{\includegraphics{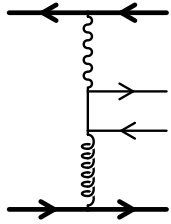}}}
\caption{\label{fig:proto}Prototype of the diagrams contributing to the photoproduction
of a $q\bar{q}$ pair in $AA$ collisions.}
\end{figure}

For this reason, it would be useful to consider a process which is
sensitive to the distinctive features of the color glass condensate
model, and yet can be observed in peripheral collisions. The second
condition implies that such a process involves electromagnetic
interactions in some way, because they are long-ranged. Since this
process should also be sensitive to the gluon density, it has to
involve quarks, which couple both to photons and gluons. Therefore,
the simplest such process is the photoproduction of $q\bar{q}$ pairs,
which has already been considered as an interesting probe for the
gluon distribution \cite{BaronB1,GreinVHS1}.  The simplest diagram for
this process has been represented in figure \ref{fig:proto}.

In this paper, we calculate this process for the peripheral collision
of two large ultra-relativistic nuclei. We treat the electromagnetic
interaction to lowest order (or, equivalently, to the leading
logarithmic approximation), and the interactions with the color field
of the nucleus to all orders in the classical color field. We find
that this process can indeed take place at large impact parameters,
and is very sensitive to the features of the gluon distribution, for
quarks whose mass is close to the saturation scale. We predict that
the transverse momentum spectrum $d\sigma_{_{T}}/dydk_\perp$ of the
produced pairs has a maximum for a momentum close to $Q_s$, which
could therefore be used as a way to measure the saturation scale.

The structure of this paper is as follows. In section \ref{sec:cgc},
we outline briefly the main aspects of the color glass condensate
model. In particular, we recall how the use of the covariant gauge can
simplify the description of the system at the classical level.

In section \ref{sec:nbar}, we detail the main steps of the calculation
of the cross-section for the photoproduction of $q\bar{q}$ pairs. In
particular, we explain how this quantity relates to Green's functions
evaluated in the classical field, how one can calculate the retarded
quark propagator in this background field, and the impact parameter
dependence of the average multiplicity.

Section \ref{sec:kt-spectrum} explains how the transverse momentum
spectrum of the produced pairs can be used as a probe of the color
glass condensate, and section \ref{sec:cross-section} deals with the
integrated cross-section.

Finally, section \ref{sec:conclusion} is devoted to concluding
remarks.

\section{The color glass condensate}
\label{sec:cgc}
A few years ago, it was suggested that the color field of a strongly
boosted nucleus can be described as a classical color field
\cite{McLerV1,McLerV2,McLerV3}, thanks to the fact that the small $x$
gluonic modes have a very large occupation number.

The classical field $A^\mu$ describing the soft modes satisfies a classical
Yang-Mills equation: 
\begin{equation}
\left[D_\mu,F^{\mu\nu}\right]=g J^\nu\; ,
\end{equation}
driven by a source $J^\nu$ generated by the hard modes. Note that this
current lives in the adjoint representation of $SU(N_c)$: $J^\nu\equiv
J^\nu_a T^a$. For a nucleus moving at very high velocity in the
positive $z$ direction, this source has the Lorentz structure\footnote{
In the following, we make extensive use of the
light-cone coordinates. For any 4-vector $x^\mu$, we define:
\begin{equation}
x^\pm\equiv {{x^0\pm x^3}\over{\sqrt{2}}}\; ,
\end{equation}
and denote ${\imb x}_\perp$ the transverse component of its 3-momentum
${\imb x}$.  With these notations, the invariant norm of $x^\mu$ is
$x^2=2x^+ x^- -{\imb x_\perp}^2$, and the scalar product of $k^\mu$
and $x^\mu$ is $k\cdot x=k^+ x^-+k^- x^+ -{\imb k_\perp}\cdot{\imb
x_\perp}$. The invariant measure $d^4x$ becomes $d^4x=dx^+dx^-d^2{\imb
x_\perp}$.  } $J^\mu=\delta^{\mu +} J^+$. Since the current $J^\mu$ is
covariantly conserved, we have:
\begin{equation}
\left[D_\mu,J^\mu\right]=0\; ,
\end{equation}
which requires\footnote{Like for $J^\nu$, we denote $\rho\equiv \rho_a T^a$.}
\begin{equation}
J^+(x^+,x^-,{\imb x}_\perp)=W(x^+,x^-,{\imb x}_\perp) 
\rho(x^-,{\imb x}_\perp)W^\dagger(x^+,x^-,{\imb x}_\perp)\; ,
\end{equation}
with
\begin{eqnarray}
&&\rho(x^-,{\imb x}_\perp)\equiv J^+(x^+_0,x^-,{\imb x}_\perp)
\; ,\nonumber\\
&&W(x^+,x^-,{\imb x}_\perp)\equiv {\rm T} \exp \left\{ig\int_{x^+_0}^{x^+}dz^+\, A^-(z^+,x^-,{\imb x}_\perp)\right\}\; .
\end{eqnarray}
  Therefore,
unlike in QED, the color source $J^+$ cannot be fixed independently of
the field $A^\mu$ itself.

However, it is possible to simplify the problem by using the
light-cone gauge defined by $A^+=0$, and by assuming that the physics
is invariant under translations along the $x^+$ axis ($\partial^-=0$)
because this is the direction of the nucleus trajectory. Under those
assumptions, one can find solutions of the Yang-Mills equation for
which $A^-=0$ (and hence $W=1$). More explicitly, the remaining
Yang-Mills equations become
\begin{eqnarray}
&&\left[D_i,F^{ij}\right]=0\; ,\nonumber\\
&&\left[D_i,\partial^+A^i\right]=-gJ^+\; .
\end{eqnarray}
The first of these equations indicates that the transverse components
are a pure gauge field. Therefore, there is a unitary matrix
$V$ such that:
\begin{equation}
A^i={i\over g} \,V^\dagger(x^-,{\imb x}_\perp)\partial^i 
V(x^-,{\imb x}_\perp)\; .
\end{equation}
By plugging this $A^i$ into the second equation, one obtains an
equation for $V$ in terms of the source $\rho$. It is however simpler
to perform a gauge transformation generated by this $V$:
\begin{equation}
A^\mu\to \widetilde
{A}^\mu\equiv {i\over g}\, V\partial^\mu V^\dagger+VA^\mu V^\dagger\; , 
\end{equation}
which leads to
\begin{eqnarray}
&&\widetilde{A}^-=0\; ,\nonumber\\
&&\widetilde{A}^i=0\; ,\nonumber\\
&&\widetilde{A}^+={i\over g}\, V\partial^+ V^\dagger\; ,
\label{eq:covariant-gauge}
\end{eqnarray}
and to the Yang-Mills equation
\begin{equation}
-\nabla_\perp^2 \widetilde{A}^+=g \widetilde{\rho}\; ,
\end{equation}
where $\widetilde{\rho}\equiv V\rho V^\dagger$ is the source in the
new gauge. This equation has the simple structure of a 2-dimensional
Poisson equation with a source $\widetilde{\rho}$, but remains very
difficult to solve in terms of the light-cone gauge source
$\rho$. However, for reasons that will become apparent later, it is
sufficient to express $V$ in terms of the covariant gauge source
$\widetilde{\rho}$:
\begin{equation}
V(x^-,{\imb x}_\perp)={\rm T}\exp\left\{ -ig^2\int_{-\infty}^{x^-}dz^- \,
{1\over{\nabla_\perp^2}}\widetilde{\rho}_a(z^-,{\imb x}_\perp)T^a\right\}\; .
\end{equation}

Having a formal solution of the classical Yang-Mills equation for a
given distribution $\widetilde{\rho}$ of the hard color sources, the
method to use this model is the following. One first expresses the
observable of interest in terms of the classical color field, or
equivalently in terms of the covariant gauge source
$\widetilde{\rho}$: ${\cal O}[\widetilde{\rho}]$. Then one must
perform an ensemble average\footnote{In a given collision, the hard
color source appears to be frozen due to time dilation. However, this
frozen configuration is going to change from event to event.}  over
the set of sources $\widetilde{\rho}$. The observed value of this
observable is therefore
\begin{equation}
\left<{\cal O}\right>\equiv \int [d\widetilde{\rho}] w[\widetilde{\rho}] {\cal O}[\widetilde{\rho}]\; .
\end{equation}
A simple ansatz for the distribution of color sources
$w[\widetilde{\rho}]$ is to take a Gaussian distribution:
\begin{equation}
w[\widetilde{\rho}]\equiv \exp\left\{- \!\!\int dx^- d^2{\imb x}_\perp
{{ \widetilde{\rho}_a(x^-,{\imb x}_\perp)\widetilde{\rho}^a(x^-,{\imb
x}_\perp)}\over{2\mu^2(x^-)}}\right\}\; .
\end{equation}
In the above formula, $\mu^2(x^-)dx^-$ has the interpretation of a
density of hard color sources per unit of transverse area, in the
slice between $x^-$ and $x^-+dx^-$. For a very ultra-relativistic
nucleus, this function is strongly peaked around $x^-=0$. At this
stage, it is easy to justify the use of either the light-cone gauge
source $\rho$ or the covariant gauge source $\widetilde{\rho}$ in
order to express observables. Indeed, the weight $w[\widetilde{\rho}]$
is invariant under the gauge transformation of the source, as is the
measure $[d\widetilde{\rho}]$. The saturation scale $Q_s^2$ is related
to the integral of $\mu^2(x^-)$ over $x^-$ (see Eq.~(\ref{eq:Qs-def})
in the appendix \ref{app:correlator} for the definition). Its value is
expected to be of the order of $1$GeV at RHIC and $2-3$GeV at LHC.

With such a weight, averages of products of sources $\widetilde{\rho}$
can all be expressed in terms of the average of the product of two of them
\begin{equation}
\left<\widetilde{\rho}_a(x^-,{\imb x}_\perp)
\widetilde{\rho}_b(y^-,{\imb y}_\perp)\right>=
\delta_{ab} \mu^2(x^-) 
\delta(x^--y^-)\delta({\imb x}_\perp-{\imb y}_\perp)\; ,
\end{equation}
thanks to Wick's theorem. More precisely, the average of the product
of an even number of $\widetilde{\rho}$'s is obtained by performing
the sum over all the possible pairings of two $\widetilde{\rho}$'s,
and the average of the product of an odd number of
$\widetilde{\rho}$'s is vanishing.

\section{Calculation of $d\sigma_{_{T}}/dy$}
\label{sec:nbar}
\subsection{Relation to Green's functions in the classical field}
Let us start by giving the connection between the average number of
pairs produced per collision in terms of a fermionic correlator in the
classical background field. For a given configuration of the hard
color sources, one can use the formalism developed in \cite{BaltzGMP1}
to express this quantity in terms of the retarded quark
propagator. Using the same notations as in \cite{BaltzGMP1}, we have
\begin{equation}
\overline{n}[\widetilde{\rho}]=\int{{d^3 {\imb q}}\over{(2\pi)^3 2\omega_{\imb q}}}
\int{{d^3 {\imb p}}\over{(2\pi)^3 2\omega_{\imb p}}}
\Big|
\overline{u}({\imb q}){\cal T}_R^{(\widetilde{\rho})}(q,-p)v({\imb p})
\Big|^2\; 
\label{eq:nbar}
\end{equation}
for the average number of pairs produced in a collision at impact
parameter ${\imb b}$. In this formula, ${\cal
T}_R^{(\widetilde{\rho})}$ denotes the interacting part of the
retarded propagator of a quark in the classical field generated by the
covariant gauge color source $\widetilde{\rho}$ and the
electromagnetic current of the other nucleus. Note also that this
quantity can be made differential with respect to the quark or the
antiquark momentum by not integrating over the corresponding
momentum. This will be useful in order to calculate the multiplicity
per unit of rapidity.

Then, averaging over the configurations $\widetilde{\rho}$
of the hard color source, one obtains $\overline{n}$:
\begin{equation}
\overline{n}=\left<\overline{n}[\widetilde{\rho}]\right>=
\int[d\widetilde{\rho}]\; w[\widetilde{\rho}]\;
\overline{n}[\widetilde{\rho}]\; .
\label{eq:nbar-avg}
\end{equation}

Note that $\overline{n}$ depends on the impact parameter ${\imb b}$ of
the collision. By integrating over impact parameter, one obtains the
total cross-section.

\subsection{Retarded quark propagator}
In order to calculate $\overline{n}$, we need the retarded propagator
of a quark moving in a classical field made of the electromagnetic
field of one nucleus, and the color field of the other nucleus. In
\cite{BaltzGMP1}, we have shown that the retarded propagator becomes
very simple in the ultra-relativistic limit. This result was derived
for QED, but since the argument was based only on causality, it can be
extended immediately to QCD. Therefore, we can write formally
(convolution products are not written explicitly) for the scattering
matrix ${\cal T}_R$:
\begin{equation}
{\cal T}_R^{(\widetilde{\rho})}=T_R^{^{QCD}}+T_R^{^{QED}}+T_R^{^{QCD}}G_R^0T_R^{^{QED}}+T_R^{^{QED}}G_R^0T_R^{^{QCD}}\; ,
\label{eq:TR}
\end{equation}
where $T_R^{^{QED}}$ and $T_R^{^{QCD}}$ are the retarded scattering
matrices of the quark in the electromagnetic field of the first
nucleus and in the color field of the second nucleus
respectively. $G_R^0$ is the free retarded propagator of a quark,
which has the following expression in Fourier space:
\begin{equation}
G_R^0(p)\equiv i {{\slp+m}\over{p^2-m^2+2p^0\epsilon}}\; .
\end{equation}
 Assuming that the nucleus acting via its electromagnetic field is
moving along the negative $z$ axis, we have in the ultra-relativistic
limit\footnote{ We denote $v_\pm\equiv(1,0,0,\pm 1)/\sqrt{2}$ the 4-velocities
of the two nuclei.}
\begin{equation}
T_R^{^{QED}}(q,p)=2\pi\delta(p^+-q^+)\slv_-\int d^2{\imb x}_\perp
\Big[e^{-ie\Lambda({\imb x}_\perp)}-1\Big] e^{i({\imb q}_\perp-{\imb p}_\perp)\cdot{\imb x}_\perp}\; ,
\label{eq:TR-QED}
\end{equation}
where $\Lambda({\imb x}_\perp)$ is the 2-dimensional Coulomb potential
of the first nucleus. Note that this potential can also be written as
\begin{equation}
\Lambda({\imb x}_\perp)=\int_{-\infty}^{+\infty}dz^- {1\over{\nabla_\perp^2}}
\rho(z^-,{\imb z}_\perp)\; ,
\label{eq:QED} 
\end{equation}
where $\rho(z^-,{\imb z}_\perp)$ is the electric charge density inside
the nucleus.

Since the second nucleus is also ultra-relativistic, we can assume that
the quark does not resolve the longitudinal structure of the color
source. The only difference with QED in that respect is the
$z^-$-ordering. If one uses the gauge defined by equations
(\ref{eq:covariant-gauge}), then the problem of the propagation of a
quark through a color glass condensate is formally similar to the QED
case, and we can generalize equation (\ref{eq:TR-QED}) without
difficulty:
\begin{equation}
T_R^{^{QCD}}(q,p)=2\pi\delta(p^--q^-)\slv_+\int d^2{\imb x}_\perp
\Big[U({\imb x}_\perp)-1\Big] e^{i({\imb q}_\perp-{\imb
p}_\perp)\cdot{\imb x}_\perp}\; ,
\end{equation}
with
\begin{equation}
U({\imb x}_\perp)\equiv {\rm T}\exp\left\{ -ig^2 \int_{-\infty}^{+\infty}dz^-{1\over{\nabla_\perp^2}}\widetilde{\rho}_a(z^-,{\imb x}_\perp)t^a\right\}\; .
\end{equation}
This formula also takes into account the fact that this nucleus is
moving in the positive $z$ direction. The inverse Laplacian acting on
the covariant gauge color source, integrated along the axis of the
collision, is the analogue of the 2-dimensional Coulomb potential in
QED (see Eq.~(\ref{eq:QED})). Note also that the covariant gauge color
source is now contracted with the generators of the fundamental
representation. More details on the eikonal approximation in a color
glass condensate can be found in \cite{KovneW1}.

For pair production, only the last two terms in Eq.~(\ref{eq:TR}) are
relevant, because one cannot produce a $q\bar{q}$ pair on-shell by
interacting with only one nucleus. After doing the integrals over the
$+$ and $-$ components of the momentum transfer (made trivial by the
delta functions), the relevant part of the retarded scattering matrix
is\footnote{We have changed $p\to -p$ as required in formula
Eq.~(\ref{eq:nbar}). In addition, we have also changed ${\imb
k}_\perp\to{\imb q}_\perp-{\imb k}_\perp$ in the first term, and
${\imb k}_\perp\to{\imb p}_\perp+{\imb k}_\perp$ in the second term
for notational convenience.}:
\begin{eqnarray}
&&{\cal T}_R^{(\widetilde{\rho})}(q,-p)=-i\int{{d^2{\imb
k}_\perp}\over{(2\pi)^2}} F_R^{^{QCD}}({\imb
k}_\perp)F_R^{^{QED}}({\imb p}_\perp+{\imb q}_\perp-{\imb
k}_\perp)\nonumber\\ 
&&\qquad\times
\Big\{{{\slv_+(\slq-\slk+m)\slv_-}\over {2q^-p^++({\imb q}_\perp-{\imb
k}_\perp)^2+m^2}} +{{\slv_-(\slk-\slp+m)\slv_+}\over {2q^+p^-+({\imb
p}_\perp-{\imb k}_\perp)^2+m^2}}\Big\}\; ,
\end{eqnarray}
where we denote
\begin{eqnarray}
&&F_R^{^{QCD}}({\imb k}_\perp)\equiv \int d^2{\imb x}_\perp 
\left[U({\imb x}_\perp)-1\right]\,e^{i{\imb k}_\perp\cdot{\imb x}_\perp}\; ,\nonumber\\
&&F_R^{^{QED}}({\imb k}_\perp)\equiv \int d^2{\imb x}_\perp 
\left[e^{-ie\Lambda({\imb x}_\perp)}-1\right]\,e^{i{\imb k}_\perp\cdot{\imb x}_\perp}\; .
\end{eqnarray}
In the following, it will prove convenient to factor out the impact
parameter dependence contained in $F_R^{^{QED}}$ (we chose the origin
of the transverse coordinates in such a way that the nucleus acting
via its color field has its center at ${\imb x}_\perp=0$, so that all
the ${\imb b}$ dependence goes into $F_R^{^{QED}}$), by writing
\begin{equation}
F_R^{^{QED}}({\imb k}_\perp)\equiv e^{i{\imb k}_\perp\cdot{\imb b}}
f_R^{^{QED}}({\imb k}_\perp)\; .
\end{equation}
Note that, at the Born level, we have
\begin{equation}
\left.f_R^{^{QED}}({\imb k}_\perp)\right|_{\rm Born}=
{{4\pi Z\alpha}\over{{\imb k}_\perp^2}}\; ,
\end{equation}
where $\alpha=e^2/4\pi$ is the electromagnetic structure constant.
The infrared sensitive denominator $1/{\imb k}_\perp^2$ produces a
large logarithm\footnote{By ``large logarithm'', we mean a logarithm
containing the center of mass energy of the collision $s$, or the
corresponding Lorentz factor $\gamma$.} in the cross-section. Coulomb
corrections beyond this Born term do not produce such a large
logarithm \cite{IvanoSS2,LeeM1,LeeM2}, and will be discarded in
this paper.

\subsection{Expression of $\overline{n}$}
Using the formula of Eq.~(\ref{eq:nbar-avg}), we can write
\begin{eqnarray}
&&\overline{n}=\int{{d^3{\imb p}}\over{(2\pi)^3 2\omega_{\imb p}}}
{{d^3{\imb q}}\over{(2\pi)^3 2\omega_{\imb q}}}
\int{{d^2{\imb k}_\perp}\over{(2\pi)^2}}
\int{{d^2{\imb l}_\perp}\over{(2\pi)^2}}
{\rm Tr}_c\left<F_R^{^{QCD}}({\imb k}_\perp)F_R^\dagger{}^{^{QCD}}({\imb l}_\perp)\right>
\nonumber\\
&&\qquad\times
F_R^{^{QED}}({\imb p}_\perp+{\imb q}_\perp-{\imb k}_\perp)
F_R^*{}^{^{QED}}({\imb p}_\perp+{\imb q}_\perp-{\imb l}_\perp)
\nonumber\\
&&\qquad\times
{\rm Tr}\,(M({\imb p},{\imb q}|{\imb k}_\perp)
M^*({\imb p},{\imb q}|{\imb l}_\perp))\; ,
\end{eqnarray}
where ${\rm Tr}_c$ is a trace over color indices, ${\rm Tr}$ is the
Dirac trace, and where we denote
\begin{eqnarray}
&&M({\imb p},{\imb q}|{\imb k}_\perp)\equiv
\overline{u}({\imb q})\Big\{
{{\slv_+(\slq-\slk+m)\slv_-}\over {2q^-p^++({\imb q}_\perp-{\imb
k}_\perp)^2+m^2}} 
\nonumber\\
&&\qquad\qquad\qquad\qquad\qquad
+{{\slv_-(\slk-\slp+m)\slv_+}\over {2q^+p^-+({\imb
p}_\perp-{\imb k}_\perp)^2+m^2}}
\Big\}v({\imb p})\; .
\end{eqnarray}
Using the results of appendix \ref{app:correlator} (see
Eqs.~(\ref{eq:app-Bs}) and (\ref{eq:app-corr-final})), we can rewrite
this as 
\begin{eqnarray}
&&\overline{n}=N_c\int{{d^3{\imb p}}\over{(2\pi)^3 2\omega_{\imb
p}}} {{d^3{\imb q}}\over{(2\pi)^3 2\omega_{\imb q}}} \int{{d^2{\imb
k}_\perp}\over{(2\pi)^2}} \int{{d^2{\imb l}_\perp}\over{(2\pi)^2}}
\nonumber\\
&&\quad\times
\int d^2{\imb x}_\perp d^2{\imb y}_\perp e^{i{\imb k}_\perp\cdot{\imb
x}_\perp} e^{-i{\imb l}_\perp\cdot{\imb y}_\perp} {\cal P}({\imb
x}_\perp){\cal P}({\imb y}_\perp)\left[1+e^{-B_2({\imb x}_\perp-{\imb y}_\perp)}-2e^{-B_1}\right]\nonumber\\
&&\quad\times F_R^{^{QED}}({\imb p}_\perp+{\imb q}_\perp-{\imb
k}_\perp) F_R^*{}^{^{QED}}({\imb p}_\perp+{\imb q}_\perp-{\imb
l}_\perp) \nonumber\\ &&\quad\times {\rm Tr}\,(M({\imb p},{\imb
q}|{\imb k}_\perp) M^*({\imb p},{\imb q}|{\imb l}_\perp))\; ,
\end{eqnarray}
where the function ${\cal P}({\imb x}_\perp)$ describes the transverse
profile of the nucleus (see the appendix \ref{app:correlator} for more
details on the effects of the nucleus finite size).  For the terms in
the square bracket which are independent of ${\imb x}_\perp$ and
${\imb y}_\perp$, the integral over those variables is trivial and
just leads to the product $\widetilde{\cal P}({\imb
k}_\perp)\widetilde{\cal P}({\imb l}_\perp)$ of the Fourier transforms
of the functions ${\cal P}$. For a large nucleus, the Fourier
transform $\widetilde{\cal P}({\imb k}_\perp)$ is very peaked around
${\imb k}_\perp=0$, with a typical width of $1/R$, where $R$ is the
radius of the nucleus.  However, it is straightforward to check that
\begin{equation}
M({\imb p},{\imb q}|0)=M({\imb p},{\imb q}|{\imb p}_\perp+{\imb q}_\perp)=0\; .
\label{eq:M-vanish}
\end{equation}
Moreover, the only scale controlling the amplitude $M({\imb p},{\imb
q}|{\imb k}_\perp)$ is the mass $m$ of the produced quark, which means
that it varies significantly only if ${\imb k}_\perp$ changes by an
amount comparable to $m$. Therefore, for heavy quarks such
that\footnote{Note that for a large nucleus of radius $R=6$fm, one has
$1/2R\sim 17$MeV.} $m\gg 1/R$, the function $M({\imb p},{\imb q}|{\imb
k}_\perp)$ remains very close to zero over all the range over which
$\widetilde{\cal P}({\imb k}_\perp)$ differs from zero. As a
consequence, we can neglect the contribution of the terms independent
of ${\imb x}_\perp$ and ${\imb y}_\perp$, and focus on the term
proportional to $\exp(-B_2({\imb x}_\perp-{\imb y}_\perp))$.

Using now the approximation of Eq.~(\ref{eq:B2-approx}) for the
function\footnote{For this approximation to be valid up to distances
$|{\imb x}_\perp-{\imb y}_\perp| \sim Q_s^{-1}$, one needs $Q_s\gg
\Lambda_{_{QCD}}$.} $B_2({\imb x}_\perp-{\imb y}_\perp)$, we see that
this function becomes much larger than $1$ as soon as $|{\imb
x}_\perp-{\imb y}_\perp|\gg Q_s^{-1}$. Since we have also
$R\gg\Lambda_{_{QCD}}^{-1}\gg Q_s^{-1}$, it is reasonable to
approximate the product ${\cal P}({\imb x}_\perp){\cal P}({\imb
y}_\perp)$ by a single factor ${\cal P}({\imb x}_\perp)$. The integral
over ${\imb x}_\perp$ and ${\imb y}_\perp$ then becomes trivial and
gives
\begin{eqnarray}
&&\overline{n}=N_c\!\!\int{{d^3{\imb p}}\over{(2\pi)^3 2\omega_{\imb
p}}} {{d^3{\imb q}}\over{(2\pi)^3 2\omega_{\imb q}}} \int\!{{d^2{\imb
k}_\perp}\over{(2\pi)^2}} \int\!{{d^2{\imb l}_\perp}\over{(2\pi)^2}}
e^{i({\imb k}_\perp-{\imb l}_\perp)\cdot{\imb b}}\widetilde{\cal P}({\imb k}_\perp-{\imb l}_\perp)
C({\imb k}_\perp)
\nonumber\\
&&\qquad\times f_R^{^{QED}}({\imb p}_\perp+{\imb q}_\perp-{\imb
k}_\perp) f_R^*{}^{^{QED}}({\imb p}_\perp+{\imb q}_\perp-{\imb
l}_\perp) \nonumber\\ &&\qquad\times {\rm Tr}\,(M({\imb p},{\imb
q}|{\imb k}_\perp) M^*({\imb p},{\imb q}|{\imb l}_\perp))\; ,
\label{eq:nbar-1}
\end{eqnarray}
where we have defined
\begin{equation}
C({\imb k}_\perp)\equiv \int d^2{\imb x}_\perp e^{i{\imb k}_\perp\cdot{\imb x}_\perp} e^{-B_2({\imb x}_\perp)}\; .
\label{eq:Ck-def}
\end{equation}
Using the Eqs.~(\ref{eq:corr}), (\ref{eq:Qs-def}), (\ref{eq:trans})
and (\ref{eq:app-Bs}) of the appendix \ref{app:correlator}, one can
see that $C({\imb k}_\perp)$ is nothing but
\begin{equation}
C({\imb k}_\perp)=\int d^2{\imb x}_\perp e^{i{\imb k}_\perp\cdot{\imb x}_\perp}
\left<U(0)U^\dagger({\imb x}_\perp) \right>\; .
\label{eq:Ck-bis}
\end{equation}
In other words, by assuming that the nucleus is large ($R\gg
\Lambda_{_{QCD}}^{-1}$), we have been able to factorize the dependence
on the nucleus size and recover the correlator
$\left<U(0)U^\dagger({\imb x}_\perp)\right>$ associated to an
hypothetical infinite nucleus.

\subsection{Impact parameter dependence}
From Eq.~(\ref{eq:nbar-1}), one can determine the main features of the
impact parameter dependence of the multiplicity $\overline{n}$.  It is
useful to recall that $M({\imb p},{\imb q}|{\imb k}_\perp)$ vanishes
when ${\imb p}_\perp+{\imb q}_\perp={\imb k}_\perp$. Moreover, since
the typical transverse momentum ${\imb p}_\perp+{\imb q}_\perp-{\imb
k}_\perp$ has to be very small for coherent photons\footnote{In order
for a photon to be coupled coherently to the total electric charge of
the nucleus, its transverse momentum should not be larger than
$1/2R$. For $R=6$fm, $1/2R\sim 17$MeV. Note that the lower bound on
the transverse momentum of the photon is $m/\gamma$, where $m$ is the
mass of the produced particle.}, compared to the other scales in
$M({\imb p},{\imb q}|{\imb k}_\perp)$ which are controlled by the
quark mass, on can use a Taylor expansion of this amplitude around the
zero:
\begin{equation}
M({\imb p},{\imb q}|{\imb k}_\perp)\approx ({\imb p}_\perp+{\imb q}_\perp-{\imb k}_\perp)\cdot {\imb L}({\imb p},{\imb q})\; ,
\end{equation}
with
\begin{eqnarray}
&&{\imb L}({\imb p},{\imb q})\equiv \overline{u}({\imb q})\left\{
{1\over{2(p^++q^+)}}\left(
{{\slv_+ {\boldsymbol \gamma}_\perp\slv_-}\over{q^-}}-
{{\slv_-{\boldsymbol \gamma}_\perp\slv_+}\over{p^-}}
\right)\right.\nonumber\\
&&\left.\qquad\qquad\qquad\qquad+{{\slv_-}\over{(p^++q^+)^2}}\left(
{{{\imb q}_\perp}\over{q^-}}-{{{\imb p}_\perp}\over{p^-}}
\right)
\right\}v({\imb p})\; .
\end{eqnarray}
In addition, the small transverse momentum of the photons allows one
to write $C({\imb k}_\perp)\approx C({\imb p}_\perp+{\imb q}_\perp)$.
Therefore, one can see that the impact parameter dependence of
$\overline{n}$ is completely controlled by the integral
\begin{equation}
I(b)\equiv\int {{d^2{\imb m}_\perp}\over{(2\pi)^2}}{{d^2{\imb
n}_\perp}\over{(2\pi)^2}}\, e^{-i({\imb m}_\perp-{\imb
n}_\perp)\cdot{\imb b}}\, \widetilde{\cal P}({\imb m}_\perp-{\imb
n}_\perp)\, {{{\imb m}_\perp\cdot{\imb L}({\imb p},{\imb
q})}\over{{\imb m}_\perp^2}}\, {{{\imb n}_\perp\cdot{\imb L}^*({\imb
p},{\imb q})}\over{{\imb n}_\perp^2}} \; ,
\end{equation}
where we have used the new variables ${\imb m}_\perp\equiv{\imb
p}_\perp+{\imb q}_\perp-{\imb k}_\perp$ and ${\imb n}_\perp\equiv{\imb
p}_\perp+{\imb q}_\perp-{\imb l}_\perp$. If the impact parameter is
very large ($b\gg R$), then the momentum difference $|{\imb
m}_\perp-{\imb n}_\perp|$ can at most be of order $1/b \ll 1/R$ because
otherwise the exponential would average to zero. On such a small
range, the function $\widetilde{\cal P}({\imb m}_\perp-{\imb
n}_\perp)$ can be replaced by $\widetilde{\cal P}(0)=\int d^2{\imb
x}_\perp {\cal P}({\imb x}_\perp)=\pi R^2$, and the integrals over
${\imb m}_\perp$ and ${\imb n}_\perp$ are trivial to perform. One
obtains:
\begin{equation}
I(b)= \pi R^2 {{{\imb b}\cdot{\imb L}({\imb p},{\imb
q})}\over{2\pi b^2}}\, {{{\imb b}\cdot{\imb L}^*({\imb
p},{\imb q})}\over{2\pi b^2}}\; .
\label{eq:large-b}
\end{equation}
Therefore, the multiplicity decreases like $1/b^2$ at large impact
parameter. If taken seriously up to $b=+\infty$, this would lead to a
logarithmic divergence in the cross-section after integration over
${\imb b}$, as is well known in problems involving long ranged
electromagnetic interactions. In reality the field does not keep its
Weizs\"acker-Williams form up to arbitrarily large transverse
distances. Indeed, at the distance $b$ from the source, the field has
a typical longitudinal extension of $b/\gamma$, where $\gamma$ is the
Lorentz factor of the electromagnetic source. This width should be
compared with the typical Compton wavelength of the particle being
produced, i.e. $1/m$ in the present case, which leads to an upper
cutoff $b=\gamma/m$ in the integral over impact parameter. This upper
bound on $b$ is equivalent to the lower bound $m/\gamma$ on the
transverse momentum of the photon.  On the lower side, one must cutoff
the integral at $b=2R$ in order to consider only peripheral
collisions. Although these cutoffs are introduced by hand, this ansatz
gives the leading logarithm correctly. This leading log approximation
is valid if there is a large range between the scales $2R$ and
$\gamma/m$. At smaller impact parameters, i.e. when $b$ is close to
$2R$, the $1/b^2$ scaling law is modified by edge effects which are
difficult to estimate analytically.

\subsection{Evaluation of $d\sigma_{_{T}}/dy$}
In order to obtain the leading logarithm of the cross-section for
peripheral collisions (the ones that are phenomenologically
interesting for photoproduction), we can use the formula
Eq.~(\ref{eq:large-b}) for all the range $2R\le b \le
\gamma/m$. Noting that
\begin{equation}
\int d^2{\imb b}\; {\imb b}^i {\imb b}^j f({\imb b}^2)={{\delta^{ij}}\over{2}}
\int d^2{\imb b}\; {\imb b}^2 f({\imb b}^2)\; , 
\end{equation}
we can write
\begin{equation}
\int d^2{\imb b}\; {{{\imb b}\cdot{\imb L}}\over{b^2}}\, {{{\imb b}\cdot{\imb L}^*}
\over{b^2}}={{{\imb L}\cdot {\imb L}^*}\over{2}}
 \;\int {{d^2{\imb b}}\over{{\imb b}^2}}\; .
\end{equation}
Therefore, we have the following expression for the total\footnote{By
``total'', we mean the cross-section obtained by counting all the
produced pairs. This quantity in general differs from the
cross-section obtained by counting only events where exactly one pair
is produced. However, when the average multiplicity is small, the two
are very close, and this distinction becomes pointless.}
cross-section for $q\bar{q}$ photoproduction in peripheral collisions:
\begin{eqnarray}
&&\sigma_{_{T}}\equiv\int\limits_{2R}^{\gamma/m}
 d^2{\imb b}\, \overline{n}=\pi R^2 {2{N_c(Z\alpha)^2}} \int\limits_{2R}^{\gamma/m}{{d^2{\imb b}}\over{{\imb b}^2}}\nonumber\\
&&\qquad\times \int{{d^3{\imb p}}\over{(2\pi)^3 2\omega_{\imb
p}}} {{d^3{\imb q}}\over{(2\pi)^3 2\omega_{\imb q}}} C({\imb p}_\perp+{\imb q}_\perp)\, {\rm Tr}\,\left(\left|{\imb L}({\imb p},{\imb q})\right|^2 \right)\; .
\label{eq:nbar-intermediate}
\end{eqnarray}
This is the formula we are going to use in the following.  In
Eq.~(\ref{eq:nbar-intermediate}), the factor which is the less well
known analytically is the correlator $C({\imb p}_\perp+{\imb
q}_\perp)$. Therefore, in the evaluation of integrals, we are going to
keep the variable ${\imb p}_\perp+{\imb q}_\perp$ (i.e. the total
transverse momentum of the pair, or approximatively the momentum
transfer between the color field and the $q\bar{q}$ pair) unintegrated
until the end. At this stage, we will have the choice to use some
analytic approximation of $C({\imb p}_\perp+{\imb q}_\perp)$ or do the
integral numerically. This approach also leaves open the possibility
to consider other saturation models.

The first step is to perform the Dirac algebra, which leads to:
\begin{eqnarray}
&&{\rm Tr}\,\left(\left|{\imb L}({\imb p},{\imb q})\right|^2
\right)={4\over{(p^-q^-(p^++q^+))^2}}
\nonumber\\
&&\qquad\qquad\times
\left\{
p^-q^-({\imb p}_\perp+{\imb q}_\perp)^2 -2{{(q^+q^-{\imb
p}_\perp+p^+p^-{\imb q}_\perp)^2}\over{(p^++q^+)^2}} \right\}\; .
\label{eq:dirac}
\end{eqnarray}
One can check that the expression of Eq.~(\ref{eq:dirac}) vanishes for
${\imb p}_\perp+{\imb q}_\perp=0$, i.e. when the momentum transfer
from the color glass condensate is zero. By using the fact that the
final state particles are on-shell, we can perform for free the
integrals over $q^+$ and $p^+$. Then, the integral over $p^-$ is very
easy because the dependence on this variable is rational. Noticing
finally that $dq^-/q^-=2 dy$, where $2y=\ln(q^+/q^-)$ is the
rapidity of the produced quark, it is very easy to obtain the
differential cross-section with respect to the rapidity\footnote{The
rapidity distribution is found to be flat here, because we do not
impose any upper bound on the energy of the quark. However, by arguing
that the produced particles cannot be more energetic than the incoming
nuclei, one of course finds that the rapidity distribution cannot
extend beyond the rapidity of the projectiles, i.e. $\pm
\ln(\gamma)$. Integration over rapidity would therefore bring another
factor $\ln(\gamma)$ in the total cross-section.} of the quark (or
equivalently, of the antiquark):

\begin{eqnarray}
&&{{d\sigma_{_{T}}}\over{dy}}=\pi R^2 {{8N_c (Z \alpha)^2}\over{\pi^2}}
\int\limits_{2R}^{\gamma/m}{{d^2{\imb b}}\over{{\imb b}^2}} \int
{{d^2{\imb p}_\perp}\over{(2\pi)^2}} {{d^2{\imb q}_\perp}\over{(2\pi)^2}}
C({\imb p}_\perp+{\imb q}_\perp)\nonumber\\
&&\qquad\qquad\qquad\times\left\{ {{({\imb p}_\perp+{\imb
q}_\perp)^2}\over{2 \omega_p^2 \omega_q^2}} -{1\over 6} \left[ {{{\imb
p}_\perp}\over{\omega_p^2}}+ {{{\imb q}_\perp}\over{\omega_q^2}}
\right]^2 \right\}\; ,
\label{eq:nbar-3}
\end{eqnarray}
with $\omega_p^2\equiv {\imb p}_\perp^2+m^2$ and $\omega_q^2\equiv
{\imb q}_\perp^2+m^2$. Let us take as one of the integration variables
the total transverse momentum of the pair ${\imb k}_\perp\equiv {\imb
p}_\perp+{\imb q}_\perp$, and keep this variable unintegrated until
the end. Then, it is easy to perform the integral over
the vector ${\imb q}_\perp$. An intermediate step is the integral over
the angle $\theta$ between ${\imb q}_\perp$ and ${\imb k}_\perp$:
\begin{eqnarray}
&&\int\limits_{0}^{2\pi}{{d\theta}\over{2\pi}}\left\{
{{{\imb k}_\perp^2}\over{2 \omega_{k-q}^2 \omega_q^2}}
-{1\over 6}
\left[
{{{\imb k}_\perp-{\imb q}_\perp}\over{\omega_{k-q}^2}}+
{{{\imb q}_\perp}\over{\omega_q^2}}
\right]^2
\right\}\nonumber\\
&&\qquad={{q_\perp^2+2k_\perp^2-m^2}\over{6(q_\perp^2+m^2)\sqrt{(m^2+q_\perp^2+k_\perp^2)^2-4q_\perp^2k_\perp^2}}}\nonumber\\
&&\qquad
-{{m^2(q_\perp^2+k_\perp^2)+(q_\perp^2-k_\perp^2)^2}\over{6\sqrt{(m^2+q_\perp^2+k_\perp^2)^2-4q_\perp^2k_\perp^2}^3}}+{{m^2}\over{6(q_\perp^2+m^2)^2}}\; .
\end{eqnarray}
Integrating then over the modulus $q_\perp$, we obtain our final
expression for $d\sigma_{_{T}}/dy$:
\begin{eqnarray}
&&{{d\sigma_{_{T}}}\over{dy}}=\pi R^2\,{{N_c (Z\alpha)^2}\over{3\pi^4}}
\int\limits_{2R}^{\gamma/m}{{d^2{\imb b}}\over{{\imb b}^2}}\nonumber\\
&&\qquad\times\int\limits_{0}^{+\infty}dk_\perp k_\perp C({\imb k}_\perp)
\left\{
1+{{4(k_\perp^2-m^2)}\over{k_\perp\sqrt{k_\perp^2+4m^2}}}{\rm arcth}\,{{k_\perp}\over{\sqrt{k_\perp^2+4m^2}}}
\right\}\; .
\label{eq:nbar-final}
\end{eqnarray}

At this stage, it may be useful to summarize all the assumptions
concerning the various parameters that entered in the derivation of
this result. First, we have assumed that the nucleus is large
\begin{equation}
R\gg \Lambda_{_{QCD}}^{-1}
\end{equation}
 in the average over the color sources (appendix \ref{app:correlator})
and before Eq.~(\ref{eq:nbar-1}). This a very good approximation for a
nucleus like gold or lead. The next assumption is that
\begin{equation}
Q_s\gg\Lambda_{_{QCD}}\; ,
\end{equation}
which has been used in the approximation that lead to
Eq.~(\ref{eq:nbar-1}). We take $\Lambda_{_{QCD}}\approx 0.2$GeV, while
$Q_s$ is expected to be of the order of $1$GeV at RHIC and $2-3$GeV
at LHC. Therefore, this assumption is only marginally satisfied at RHIC,
and probably much better at LHC. We have also assumed that the quark
mass is large enough
\begin{equation}
m\gg R^{-1}
\end{equation}
before Eqs.~(\ref{eq:nbar-1}) and (\ref{eq:nbar-intermediate}). In
addition, for the leading log approximation in the integral over
impact parameter to be justified, one needs a large enough center of
mass energy
\begin{equation}
\gamma \gg mR\; .
\end{equation}

\section{$k_\perp$ spectrum}
\label{sec:kt-spectrum}
\begin{figure}[ht]
\centerline{
\resizebox*{!}{7.5cm}{\includegraphics{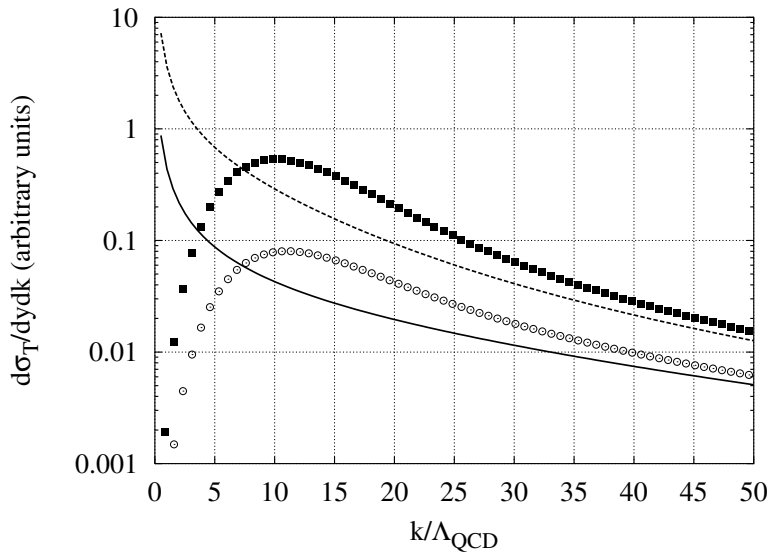}}}
\caption{\label{fig:spectrum}Plot of $d\sigma_{_{T}}/dy dk_\perp$ as a
function of $k_\perp/\Lambda_{_{QCD}}$, in arbitrary units (only the
$k_\perp$ dependent factors of Eq.~(\ref{eq:kt-spectrum}) have been
plotted). The value of $Q_s/\Lambda_{_{QCD}}$ is set to $10$. The
dotted curves are based on a numerical evaluation of the full $C({\imb
k}_\perp)$ as predicted in the color glass condensate model. The
continuous curves have been obtained with the asymptotic form $C({\imb
k}_\perp)=2Q_s^2/k_\perp^4$, which is the lowest order in
$Q_s^2$. Open dots and solid curve: $m/\Lambda_{_{QCD}}=23$ (bottom),
filled dots and dashed curve: $m/\Lambda_{_{QCD}}=8$ (charm), with
$\Lambda_{_{QCD}}\approx 0.2$GeV. }
\end{figure}

In Eq.~(\ref{eq:nbar-final}), $d\sigma_{_{T}}/dy$ is given by a
1-dimensional integral involving the correlator $C(k_\perp)$ defined
in Eqs.~(\ref{eq:Ck-def}) and (\ref{eq:Ck-bis}). If all one needs is a
spectrum as a function of the total momentum of the pair, then it is
not necessary to integrate this expression further, and the $k_\perp$
spectrum is directly given by
\begin{eqnarray}
&&{{d\sigma_{_{T}}}\over{dydk_\perp}}=\pi R^2\,{{2N_c (Z\alpha)^2}\over{3\pi^3}}\ln\left({{\gamma}\over{2mR}}\right)\nonumber\\
&&\qquad\times
 k_\perp C({\imb k}_\perp)
\left\{
1+{{4(k_\perp^2-m^2)}\over{k_\perp\sqrt{k_\perp^2+4m^2}}}{\rm arcth}\,{{k_\perp}\over{\sqrt{k_\perp^2+4m^2}}}
\right\}
\; .
\label{eq:kt-spectrum}
\end{eqnarray}
We observe that this spectrum has the interesting property of being
directly proportional to the correlator $C({\imb k}_\perp)$. According
to the discussion of appendix \ref{app:Ck}, this correlator is
sensitive to saturation effects as soon as $k_\perp \lesssim
Q_s$. The possibility to observe saturation effects by measuring the
$k_\perp$ spectrum of $q\bar{q}$ pairs produced via photoproduction is
illustrated in the figure \ref{fig:spectrum}, which contains plots of
the $k_\perp$ spectrum for the charm and bottom quarks.
One can see a qualitatively important difference between the
prediction based on the color glass condensate (dots) and the result
one obtains by assuming no saturation (solid curves), i.e. by using
only the first order term in $Q_s^2$ for $C(k_\perp)$:
\begin{equation}
C(k_\perp)\approx 2{{Q_s^2}\over{k_\perp^4}}\; ,
\label{eq:Ck-asymp}
\end{equation}
derived in appendix \ref{app:Ck}. The color glass condensate model
 predicts a maximum in the $k_\perp$ spectrum, for a value of
 $k_\perp$ closely related to the saturation scale $Q_s$. It is in
 fact easy to understand why, for a large mass $m$, the location of
 the maximum is completely controlled by $Q_s$. Indeed, if the mass is
 very large, the function that appears inside the curly brackets in
 Eq.~(\ref{eq:kt-spectrum}) can be approximated by:
\begin{equation}
\left.
1+{{4(k_\perp^2-m^2)}\over{k_\perp\sqrt{k_\perp^2+4m^2}}}{\rm arcth}\,{{k_\perp}\over{\sqrt{k_\perp^2+4m^2}}}\right|_{k_\perp\ll m}
={7\over 6}{{k_\perp^2}\over{m^2}}+{\cal O}\left({{k_\perp^4}\over{m^4}}\right)\; .
\label{eq:bracket-small-k}
\end{equation}
This indicates that the mass dependence becomes a trivial factor
$m^{-2}$ which does not affect the location of the maximum. The
location of the maximum is entirely determined by the function
$k_\perp^3 C(k_\perp)$, and therefore depends only on $Q_s$. A
semi-analytic estimate of the value of $k_\perp$ at the maximum leads
to
\begin{equation}
k_\perp^{\rm max}\approx Q_s 
\sqrt{{3\over{2\pi}}\ln\left({{Q_s}\over{\Lambda_{_{QCD}}}}\right)}\; .
\end{equation}
For
intermediate masses, this is not exactly true, but we have checked
numerically that the location of the maximum varies very little. For
instance, for $Q_s/\Lambda_{_{QCD}}=10$, the maximum varies from
$k_\perp/\Lambda_{_{QCD}}=9$ to $k_\perp/\Lambda_{_{QCD}}=12$ if one
varies the mass between $m/\Lambda_{_{QCD}}=5$ and
$m/\Lambda_{_{QCD}}=100$. Observing experimentally this maximum could
therefore be an evidence for saturation, and provide a measurement of
the saturation scale.

\section{Integrated cross-section}
\label{sec:cross-section}
In this section, we discuss various properties of the cross-section
integrated over the transverse momentum of the pair. In particular, we
provide analytical results for the limit of large mass ($m\gg Q_s$)
and the limit of small mass ($m\ll Q_s$). To that effect, we will need
also the large $k_\perp$ expression of the function that appears
inside the curly brackets in Eq.~(\ref{eq:nbar-final}):
\begin{equation}
\left.1+{{4(k_\perp^2-m^2)}\over{k_\perp\sqrt{k_\perp^2+4m^2}}}{\rm arcth}\,{{k_\perp}\over{\sqrt{k_\perp^2+4m^2}}}\right|_{k_\perp\gg m}\approx
4\ln(k_\perp/m)\; .
\label{eq:bracket-large-k}
\end{equation}

\subsection{Large mass limit}
In the limit of large quark mass, the typical $k_\perp$ in the
integral of Eq.~(\ref{eq:nbar-final}) is of order $m$ or smaller. In
addition, since $m\gg Q_s$, then so will be $k_\perp$. For such large
values of $k_\perp$, we can use Eq.~(\ref{eq:Ck-asymp}) for
$C(k_\perp)$.  Note also that $C(k_\perp)$ is strongly suppressed
compared to this asymptotic value if $k_\perp\lesssim
\Lambda_{_{QCD}}$ (see figure \ref{fig:Ck} in appendix \ref{app:Ck}),
which provides a natural lower bound for the integral over $k_\perp$.

Making use of Eq.~(\ref{eq:bracket-small-k}) and performing the
integral over $k_\perp$, we obtain the following asymptotic expression
for $d\sigma_{_{T}}/dy$ when $m\gg Q_s$:
\begin{equation}
\left.{{d\sigma_{_{T}}}\over{dy}}\right|_{m\gg Q_s} =\pi R^2\,{{14N_c (Z
\alpha)^2}\over{9\pi^3}}
{{Q_s^2}\over{m^2}}\ln\left({m\over{\Lambda_{_{QCD}}}}\right)
\ln\left({{\gamma}\over{2mR}}\right)
\;
.
\label{eq:sigma-large-m}
\end{equation}

Note that this formula cannot be used for very large masses, such that
the argument $\gamma/2mR$ of the logarithm becomes close to $1$ (or
smaller). Indeed, the logarithmic approximation used in this paper is
no longer valid. Moreover, the kinematical lower bound $m/\gamma$ for
the transverse momentum of the photon being larger than the upper
limit $1/2R$ of the Weizs\"acker-Williams photon spectrum, the photons
that can produce such a massive pair of particles has to come from the
tail of the photon spectrum, which is exponentially
suppressed. Details on this can be found in \cite{KleinNV1} where is
studied the photoproduction of top quark pairs in peripheral heavy ion
collisions.

\subsection{Small mass limit}
Similarly, one can deal analytically with the case of a very small
mass ($m\ll Q_s$). For this case, it turns out to
be simpler to revert the integral in Eq.~(\ref{eq:nbar-final}) to
transverse coordinate space\footnote{Because the expression between
the curly brackets vanishes at $k_\perp=0$, it is possible to replace
$e^{-B_2}$ by $e^{-B_2}-1$ in the definition of the correlator
$C(k_\perp)$. We have made use of this freedom here.}
\begin{equation}
{{d\sigma_{_{T}}}\over{dy}}=\pi R^2\,{{2N_c (Z \alpha)^2}\over{3\pi^3}}
\ln\left({\gamma\over{2mR}}\right)
\,4\pi^2\int\limits_{0}^{m^{-1}}dx_\perp\, x_\perp \left(1-e^{-B_2(x_\perp)}\right)
\left({{2}\over{\pi x_\perp^2}}\right)\; ,
\end{equation}
where the last factor is nothing but the inverse Fourier transform of
Eq.~(\ref{eq:bracket-large-k}), in the range $x_\perp\ll 1/m$. For $x_\perp\gg
1/m$, on can check that this Fourier transform is strongly suppressed, hence
the phenomenological upper bound in the integral over $x_\perp$.
Because of the behavior of the function $B_2({\imb x}_\perp)$, the
combination $1-\exp(-B_2({\imb x}_\perp))$ behaves mostly like
$\theta(x_\perp-1/Q_s)$ where $\theta$ is the Heaviside step
function. Therefore, we find for the small mass limit:
\begin{equation}
\left.{{d\sigma_{_{T}}}\over{dy}}\right|_{m\ll Q_s} =\pi R^2\,{{16N_c (Z
\alpha)^2}\over{3\pi^2}}
\ln\left({{Q_s}\over{m}}\right)\ln\left({\gamma\over{2mR}}\right)\;
.
\label{eq:sigma-small-m}
\end{equation}

One can note that the integrated cross-section could also have been
obtained from the expression of the structure function $F_2(x,Q^2)$
derived in \cite{McLerV4}. Taking the large and small mass limit in
$F_2$, and convoluting with the Weizs\"acker-Williams spectrum, one
can recover the Eqs.~(\ref{eq:sigma-large-m}) and
(\ref{eq:sigma-small-m}).

\subsection{Numerical results}
\begin{figure}[ht]
\resizebox*{!}{7.5cm}{\includegraphics{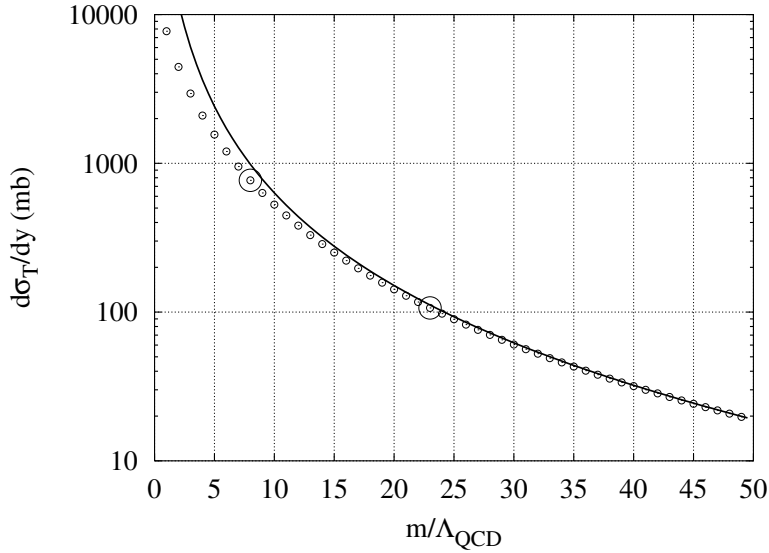}}
\caption{\label{fig:mass-plot}Plot of $d\sigma_{_{T}}/dy$ as a function of
$m/\Lambda_{_{QCD}}$. In this plot, $Q_s/\Lambda_{_{QCD}}=10$. We use
$R=6$fm and $\gamma=3000$. The open dots are a numerical evaluation of
the integral in Eq.~(\ref{eq:nbar-final}), while the solid curve is
the asymptotic formula of Eq.~(\ref{eq:sigma-large-m}), in which we
have replaced $\ln(m/\Lambda_{_{QCD}})$ by
$\ln(m/\Lambda_{_{QCD}})+0.92$ (the additive constant has been
adjusted so that it fits at large values of $m$). We have circled the
points corresponding to the mass of the charm and bottom quarks,
assuming $\Lambda_{_{QCD}}\approx 0.2$GeV.}
\end{figure}

One can also evaluate numerically the integral over transverse
momentum in Eq.~(\ref{eq:nbar-final}). The result of this calculation
is plotted in figure \ref{fig:mass-plot}. One can see that the
asymptotic formula obtained in Eq.~(\ref{eq:sigma-large-m}) is
reasonably accurate for $m \ge Q_s$, up to an additive constant
correction to the $\ln(m/\Lambda_{_{QCD}})$ which could not be
determined analytically and has been fitted.

In addition, one can see that for large nuclei ($R=6$fm) at LHC
energies ($\gamma=3000$), the value of the cross-section is about
$800$mb for charm quarks, and $100$mb for bottom quarks. This is before
any experimental cut is applied on the range of $k_\perp$.

\section{Conclusion}
\label{sec:conclusion}
In this paper, we have evaluated the cross-section for the
photoproduction of $q\bar{q}$ pairs in the peripheral collision of two
ultra-relativistic heavy nuclei. Both nuclei are treated as
classical sources, respectively of electromagnetic field and color
field. The classical color field is described using the color glass
condensate model.

It appears that this process is very sensitive to the properties of
the gluon distribution in the region where saturation might play an
important role.  In particular, we find that the saturation of the
gluon density modifies significantly the shape of the $k_\perp$
spectrum of the produced pairs. We predict a maximum in
$d\sigma_{_{T}}/dy dk_\perp$ when $k_\perp$ is comparable to the
saturation scale $Q_s$. Therefore, one could possibly use this process
as a test of the color glass condensate model, and as a way to
measure the saturation scale.

It is also interesting to study the case of the diffractive production
of $q\bar{q}$ pairs, where the exchange between the color source and
the pair is color singlet and carries no net transverse momentum. One
would then have events where the quark and antiquark have opposite
transverse momenta, with no other object in the final state. This
calculation is a work in progress.

\vskip 2mm
\noindent{\bf Acknowledgments:} This work is supported by DOE under
grant DE-AC02-98CH10886.  A.P. is supported in part by the
A.-v.-Humboldt foundation (Feo\-dor-Lynen program). We would like to
thank L. McLerran for insightful comments on this work, as well as
A. Dumitru, S. Gupta, J. Jalilian-Marian, D. Kharzeev, V. Serbo and
R. Venugopalan for discussions on related topics.

\appendix

\section{Calculation of $\left<U({\imb x}_\perp)U^\dagger({\imb y}_\perp)\right>$}
\label{app:correlator}
\subsection{$\left<U({\imb x}_\perp)\right>$}
In this appendix, we calculate the correlator $\left<U({\imb
x}_\perp)U^\dagger({\imb y})\right>$ with
\begin{equation}
U({\imb x}_\perp)={\rm T}\exp\left\{ -ig^2 \int_{-\infty}^{+\infty} dz^- {1\over{\nabla_\perp^2}} \widetilde{\rho}_a(z^-,{\imb x}_\perp)t^a\right\}\; ,
\end{equation}
where $t^a$ is a color matrix in the fundamental representation of
$SU(N_c)$. More explicitly:
\begin{equation}
{1\over{\nabla_\perp^2}} \widetilde{\rho}_a(z^-,{\imb x}_\perp)=\int d^2{\imb z}_\perp G_0({\imb x}_\perp-{\imb z}_\perp) \widetilde{\rho}_a(z^-,{\imb z}_\perp)\; ,
\end{equation}
where $G_0$ is the propagator associated to the 2-dimensional
Laplacian. This propagator satisfies:
\begin{equation}
{{\partial^2}\over{\partial {\imb z}_\perp^2}} G_0({\imb
x}_\perp-{\imb z}_\perp)=\delta({\imb x}_\perp-{\imb z}_\perp)\; .
\end{equation}
More explicitly, we have
\begin{equation}
G_0({\imb x}_\perp-{\imb z}_\perp)=-\int {{d^2{\imb k}_\perp}\over{(2\pi)^2}}
{{e^{i{\imb k}_\perp\cdot({\imb x}_\perp-{\imb z}_\perp)}}\over{{\imb k}_\perp^2}}\; .
\label{eq:prop-FT}
\end{equation}

Let us first consider the average of $U({\imb x}_\perp)$. It
is in fact useful to calculate the slightly more general quantity
$\left<U(a^-,b^-|{\imb x}_\perp)\right>$, where we define
\begin{equation}
U(a^-,b^-|{\imb x}_\perp)={\rm T}\exp\left\{ -ig^2 \int_{a^-}^{b^-} dz^-
 {1\over{\nabla_\perp^2}} \widetilde{\rho}_a(z^-,{\imb x}_\perp)t^a\right\}\;
 .
\end{equation}
By expanding the time-ordered exponential, we have
\begin{eqnarray}
&&\!\!\!\!\!\!\left<U(a^-,b^-|{\imb
x}_\perp)\right>=\sum_{n=0}^{+\infty}{{(-ig^2)^n}\over{n!}}  \int
\prod_{i=1}^{n}\left[
d^2{\imb z}_{i\perp}G_0({\imb x}_\perp-{\imb z}_{i\perp})\right]\nonumber\\
&&\times
\int_{a^-}^{b^-}\!\!\!dz_1^- \int_{z_1^-}^{b^-}\!\!\!dz_2^-\cdots
\int_{z_{n-1}^-}^{b^-}\!\!\!dz_n^-
\left<\widetilde{\rho}_{a_1}(z_1^-,{\imb z}_{1\perp})\cdots
\widetilde{\rho}_{a_n}(z_n^-,{\imb z}_{n\perp})\right> t^{a_1}\cdots
t^{a_n}\; .\nonumber\\
&&
\end{eqnarray}
The bracket under the integral can be reduced to averages of products
of two $\widetilde{\rho}$'s thanks to Wick's theorem.  To make the
discussion more visual, we represent the Wilson line integral in $U$ by a
bold straight line, the free 2-dimensional propagators $G_0$ by  wavy lines, and the elementary correlator $\left<\widetilde{\rho}_a \widetilde{\rho}_b\right>$ by a cross inserted between two propagators:
\setbox1=\hbox to 3cm{\hfill\resizebox*{3cm}{!}{\includegraphics{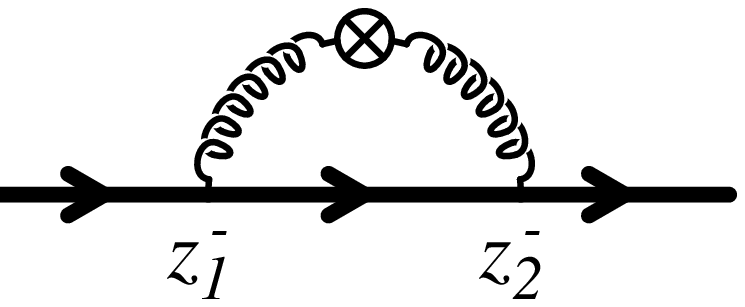}}}
\begin{equation}
\raise -7mm\box1\equiv G_0({\imb x}_\perp-{\imb z}_{1\perp})G_0({\imb x}_\perp-{\imb z}_{2\perp})\left<\widetilde{\rho}_{a_1}(z_1^-,{\imb z}_{1\perp})
\widetilde{\rho}_{a_2}(z_2^-,{\imb z}_{2\perp})\right>\; .
\end{equation}
Because of the $\delta(z_1^--z_2^-)$ contained in such a contraction,
we will call it a ``tadpole'' in the following.  The only possibility
for multiple pairings on a line ordered in $z^-$ is to have adjacent
tadpoles, like in the following diagram: \setbox1=\hbox to
7cm{\hfill\resizebox*{7cm}{!}{\includegraphics{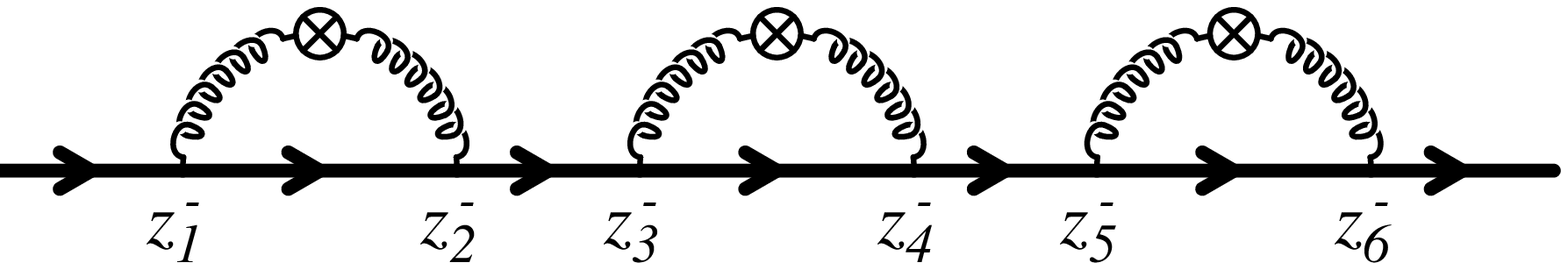}}}
\begin{displaymath}
\raise -3mm\box1\; .
\end{displaymath}
Diagrams with nested or overlapping loops are vanishing because they
have no support on the $z^-$ axis. Noticing also that
\begin{equation}
\int_{z_1^-}^{b^-}dz_2^- \delta(z_1^--z_2^-)={1\over 2}\; ,
\end{equation}
we finally obtain
\begin{equation}
\left<U(a^-,b^-|{\imb
x}_\perp)\right>=\exp \left\{-{{g^4}\over 2} (t^at_a)
\left[\int_{a^-}^{b^-}dz^-\mu^2(z^-)\right]
\int d^2{\imb z}_\perp G_0^2({\imb x}_\perp-{\imb z}_\perp)\right\}\; .
\end{equation}
We can recognize in this expression the color density per unit of
transverse area obtained by integrating $\mu^2$ along the $z^-$ axis between
$a^-$ and $b^-$.

\subsection{$\left<U({\imb x}_\perp)U^\dagger({\imb y}_\perp)\right>$}
Let us now turn to the case of the average of the product $U({\imb
x}_\perp) U^\dagger({\imb y}_\perp)$. One can have pairings that
connect the Wilson lines corresponding to the two $U$'s, but the $z^-$
ordering prevents the corresponding links to be crossed. We therefore
have only ``ladder'' topologies, or ``tadpole'' topologies similar to
those encountered above. In summary, we need to resum all the
topologies like: \setbox1=\hbox to
5cm{\hfill\resizebox*{5cm}{!}{\includegraphics{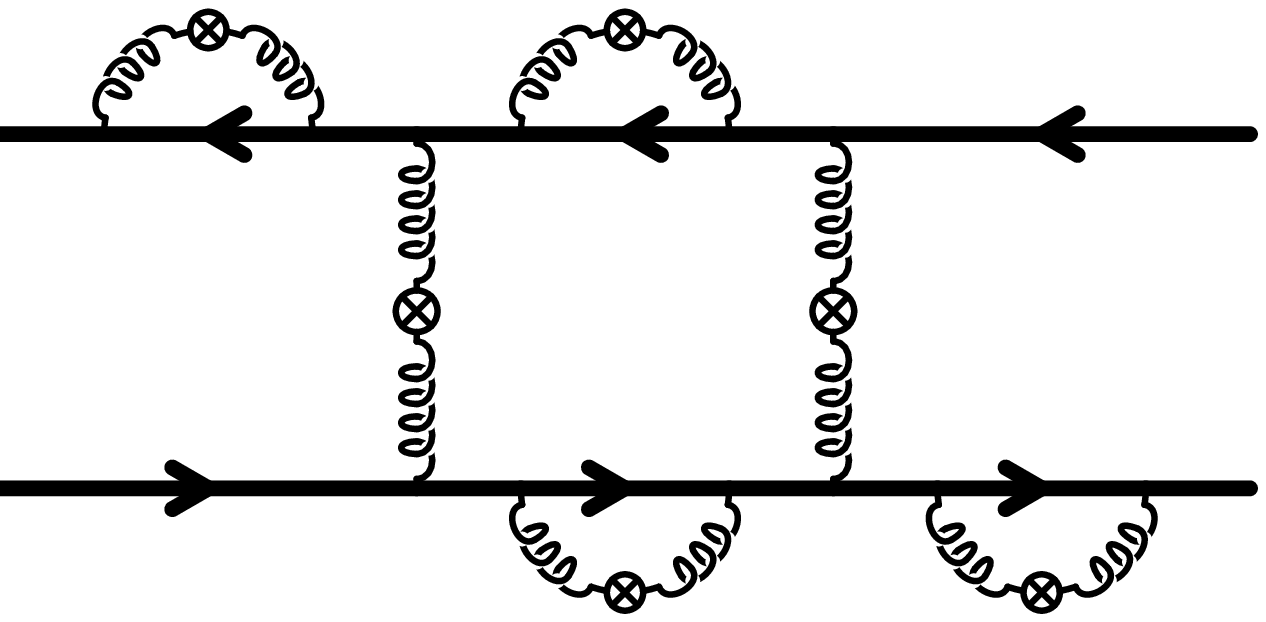}}}
\begin{displaymath}
\raise -3mm\box1\; .
\end{displaymath}
Diagrams that do not fall in this class vanish because of the
ordering. In order to evaluate the correlator
$\left<UU^\dagger\right>$, we proceed as above, noticing that for a link
connecting the two Wilson lines, the integral over $z^-$ does not
produce a factor $1/2$ because there is no relative order between the
$z^-$'s on the two lines. Then, we can write the correlator as a sum
over the number $n$ of rungs in the ladder:
\begin{eqnarray}
&&\left<U(a^-,b^-|{\imb x}_\perp) U^\dagger(a^-,b^-|{\imb y}_\perp)\right>
=\sum_{n=0}^{+\infty}
\int_{a^-}^{b^-}dz_1^- \int_{z_1^-}^{b^-}dz_2^-\cdots\int_{z_{n-1}^-}^{b^-}dz_n^-\nonumber\\
&&\qquad\times
\left<U(a^-,z_1^-|{\imb x}_\perp)\right>
\left<U(z_1^-,z_2^-|{\imb x}_\perp)\right>
\cdots
\left<U(z_n^-,b^-|{\imb x}_\perp)\right>\nonumber\\
&&\qquad\times
\left<U(a^-,z_1^-|{\imb y}_\perp)\right>
\left<U(z_1^-,z_2^-|{\imb y}_\perp)\right>
\cdots
\left<U(z_n^-,b^-|{\imb y}_\perp)\right>
\nonumber\\
&&\qquad\times
\left[g^4 (t^at_a)\int d^2{\imb z}_\perp G_0({\imb x}_\perp-{\imb
z}_\perp)G_0({\imb y}_\perp-{\imb z}_\perp)\right]^n
\mu^2(z_1^-)\cdots\mu^2(z_n^-)\; ,\nonumber\\
&&
\end{eqnarray}
where we have already resummed all the tadpole insertions on the
Wilson lines between the rungs of the ladder. One can notice that all
the color matrices come in pairs $t^at_a$, which are proportional to
the unit matrix.  Due to their exponential structure, the tadpole
insertions can be combined into a single exponential which depends
only on the endpoints $a^-$ and $b^-$, and can therefore be factored
out of the integral. The sum over $n$ then trivially exponentiates, so
that:
\begin{eqnarray}
&&\!\!\!\!\!\!\left<U(a^-,b^-|{\imb x}_\perp) U^\dagger(a^-,b^-|{\imb y}_\perp)\right>
=\left<U(a^-,b^-|{\imb x}_\perp)\right>\left<U(a^-,b^-|{\imb y}_\perp)\right>\nonumber\\
&&\;\times
\exp \left\{g^4 (t^at_a)
\left[\int_{a^-}^{b^-}dz^-\mu^2(z^-)\right]
\int d^2{\imb z}_\perp G_0({\imb x}_\perp-{\imb z}_\perp)G_0({\imb y}_\perp-{\imb z}_\perp)\right\}\; .\nonumber\\
&&
\end{eqnarray}
Combining all the factors, taking $a^-\to-\infty$, $b^-\to+\infty$, we
can write:
\begin{eqnarray}
&&\left<U({\imb x}_\perp)U^\dagger({\imb y}_\perp)\right>
\nonumber\\
&&
=\!
\exp \!\left\{\!-{{g^4}\over 2}(t^at_a)\!
\left[\int\limits_{-\infty}^{+\infty}\!\!\!dz^-\mu^2(z^-)\right]
\!\!\int \!\!d^2{\imb z}_\perp \left[G_0({\imb x}_\perp\!-\!{\imb z}_\perp)-G_0({\imb y}_\perp\!-\!{\imb z}_\perp)\right]^2\!\right\} .\nonumber\\
&&
\label{eq:corr}
\end{eqnarray}
One usually introduces here the saturation scale $Q_s$, defined by\footnote{
We have chosen here a definition of $Q_s^2$ which is particularly convenient when coupling the
color field to fermions, because it absorbs the Casimir in the fundamental representation. Other
conventions have been used in the literature. For instance, \cite{larry} defines
\begin{equation}
Q_s^2=2\pi N_c \alpha_s^2 \int\limits_{-\infty}^{+\infty}dz^-\mu^2(z^-)\; .
\end{equation}
}
\begin{equation}
Q_s^2\equiv {{g^4}\over 2}(t^at_a)\!
\int\limits_{-\infty}^{+\infty}\!\!dz^-\mu^2(z^-)\; .
\label{eq:Qs-def}
\end{equation}

\subsection{Finite size effects}
The above calculation has been performed under the assumption that the
color source has an infinite transverse extension, which leads to
translation invariance in the transverse plane for the correlator:
\begin{equation}
\left<U({\imb x}_\perp)U^\dagger({\imb y}_\perp)\right>=
\left<U(0)U^\dagger({\imb y}_\perp-{\imb x}_\perp)\right>\; .
\label{eq:trans}
\end{equation}
In order to see how the finite size of the nucleus enters in this
correlator, we have to go back and replace $\mu^2(z^-)$ by
$\mu^2(z^-){\cal P}({\imb z}_\perp)$, where the function ${\cal
P}({\imb z}_\perp)$ describes the transverse profile of the nucleus
(if one neglects edge effects, this function is 1 inside the nucleus
and 0 outside). Because of our ansatz to factorize\footnote{This is of
course not exact because for a spherical nucleus the transverse
profile depends on $z^-$ (or, equivalently, $Q_s$ is smaller at the
edges of the nucleus than at the center), but this ansatz captures the
main features due to the nucleus finite size. Indeed, in this
particular problem, the photon that hits the nucleus has a very small
transverse momentum and cannot resolve the transverse size of the
nucleus. This photon therefore sees a density of color charges
averaged over the transverse section of the nucleus, or an averaged
saturation scale $Q_s$. This averaged $Q_s$ is dominated by the rather
large value at the center of the nucleus.} the profile from the
density $\mu^2(z^-)$, the transverse profile does not change the
calculation of the correlator. Simply, the argument of the exponential
of Eq.~(\ref{eq:corr}) becomes (up to a sign):
\begin{equation}
A({\imb x}_\perp,{\imb y}_\perp)\equiv Q_s^2\int d^2{\imb z}_\perp {\cal
P}({\imb z}_\perp) \Big[G_0({\imb x}_\perp-{\imb z}_\perp)-G_0({\imb
y}_\perp-{\imb z}_\perp)\Big]^2\; .
\end{equation}
We see that the finite size of the nucleus reduces the support of this
integral to the transverse section of the nucleus. Inspection of this
integral indicates that there is a residual logarithmic infrared
divergence (a stronger, quadratic, divergence is cancelled thanks to
the difference of the two propagators). This divergence is in fact
well known and it has been argued that it is naturally screened by
confinement (a feature which is not present in the color glass
condensate model) at a scale comparable to the nucleon size (see
\cite{LamM1} for a more elaborate discussion of this issue), which is
of the order of $\Lambda_{_{QCD}}^{-1}$. Practically, one can achieve
this by cutting off the propagator $G_0({\imb x}_\perp-{\imb
z}_\perp)$ in order to make it vanish for distances larger than
$\Lambda_{_{QCD}}^{-1}$. For instance, one can regularize the Fourier
transform in Eq.~(\ref{eq:prop-FT}) by replacing ${\imb k}_\perp^2$ by
${\imb k}_\perp^2+\Lambda_{_{QCD}}^2$, which leads to a propagator
proportional to $K_0(\Lambda_{_{QCD}}({\imb x}_\perp-{\imb
z}_\perp))$. All the predictions made by using this prescription will
depend on $\Lambda_{_{QCD}}^{-1}$ in a universal way
(i.e. independently of the details of the implementation of the
cutoff) as long as one looks only at momenta large compared to
$\Lambda_{_{QCD}}$.

Under the assumption that the nucleus is very large (i.e. its radius
$R$ satisfies $R\gg \Lambda_{_{QCD}}^{-1}$), one can split the
function $A({\imb x}_\perp,{\imb y}_\perp)$ according to whether
${\imb x}_\perp$ and ${\imb y}_\perp$ are inside or outside of the
nucleus:
\begin{eqnarray}
A({\imb x}_\perp,{\imb y}_\perp)=&&(1-{\cal P}({\imb x}_\perp))(1-{\cal P}({\imb y}_\perp)) \times 0\nonumber\\
&&+ {\cal P}({\imb x}_\perp)(1-{\cal P}({\imb y}_\perp)) B_1({\imb x}_\perp)\nonumber\\
&&+(1-{\cal P}({\imb x}_\perp)){\cal P}({\imb y}_\perp) B_1({\imb y}_\perp)\nonumber\\
&&+{\cal P}({\imb x}_\perp){\cal P}({\imb y}_\perp) B_2({\imb x}_\perp-{\imb y}_\perp)\; ,
\label{eq:exponent}
\end{eqnarray}
where we denote
\begin{eqnarray}
&&B_1({\imb x}_\perp)\equiv Q_s^2 \int d^2{\imb z}_\perp G_0^2({\imb x}_\perp-{\imb z}_\perp)\; ,\nonumber\\
&&B_2({\imb x}_\perp)\equiv Q_s^2 \int d^2{\imb z}_\perp \left[G_0({\imb z}_\perp)-G_0({\imb z}_\perp-{\imb x}_\perp)\right]^2\; .
\label{eq:app-Bs}
\end{eqnarray}
In Eq.~(\ref{eq:exponent}), the neglected terms are suppressed by
powers of $(R\Lambda_{_{QCD}})^{-1}$. Using now the Fourier transform
representation of $G_0$, it is trivial to arrive at
\begin{equation}
B_2({\imb x}_\perp)={{Q_s^2}\over \pi}\int\limits_{0}^{+\infty}{{dp}\over{p^3}}(1-J_0(px_\perp))\; .
\end{equation}
Cutting off the logarithmic divergence by $\Lambda_{_{QCD}}$, we find
\begin{equation}
B_2({\imb x}_\perp)\approx {{Q_s^2x_\perp^2}\over{4\pi}}\ln\left({1\over{x_\perp\Lambda_{_{QCD}}}}\right)\; ,
\label{eq:B2-approx}
\end{equation}
an approximation valid for transverse separations ${\imb x}_\perp$
much smaller than $\Lambda_{_{QCD}}^{-1}$. For $B_1$, we find to
lowest order in $\Lambda_{_{QCD}}/Q_s$:
\begin{equation}
B_1 \sim {{Q_s^2}\over{\Lambda_{_{QCD}}^2}}\gg 1\; .
\end{equation}

Using those results, we can rewrite the two correlators considered in
this appendix as:
\begin{equation}
\left<U({\imb x}_\perp)\right>=\left<U^\dagger({\imb x}_\perp)\right>=
1-{\cal P}({\imb x}_\perp)+{\cal P}({\imb x}_\perp)e^{-B_1}\; ,
\end{equation}
and
\begin{eqnarray}
\left<U({\imb x}_\perp)U^\dagger({\imb y}_\perp)\right>=&&
(1-{\cal P}({\imb x}_\perp))(1-{\cal P}({\imb y}_\perp))\nonumber\\
&&+ {\cal P}({\imb x}_\perp)(1-{\cal P}({\imb y}_\perp)) e^{-B_1}\nonumber\\
&&+(1-{\cal P}({\imb x}_\perp)){\cal P}({\imb y}_\perp) e^{-B_1}\nonumber\\
&&+{\cal P}({\imb x}_\perp){\cal P}({\imb y}_\perp) e^{-B_2({\imb x}_\perp-{\imb y}_\perp)}\; .
\end{eqnarray}
From this, it is trivial to obtain the correlator needed in section
\ref{sec:nbar}
\begin{equation}
\left<(U({\imb x}_\perp)-1)(U^\dagger({\imb y}_\perp)-1)\right>= {\cal
P}({\imb x}_\perp){\cal P}({\imb y}_\perp)\left[1+e^{-B_2({\imb
x}_\perp-{\imb y}_\perp)}-2 e^{-B_1}\right]\; .
\label{eq:app-corr-final}
\end{equation}
Note that in the derivation of $\overline{n}$, the factor $U({\imb
x}_\perp)-1$ comes from the retarded amplitude ${\cal T}_R$ while the
factor $U^\dagger({\imb y}_\perp)-1$ comes from its complex
conjugate. The coordinates ${\imb x}_\perp$ and ${\imb y}_\perp$ can
be seen as the transverse coordinate of either the quark or the
antiquark. Therefore, the proportionality of this correlator to the
factor ${\cal P}({\imb x}_\perp){\cal P}({\imb y}_\perp)$ simply means
that in order to produce a $q\bar{q}$ pair, either the quark or the
antiquark has to intercept the nucleus that acts via its color field
(and that, both in the amplitude and in its complex conjugate).

\section{Asymptotic expansion of $C(k_\perp)$}
\label{app:Ck}
The pair multiplicity depends crucially on the correlator defined by
\begin{equation}
C(k_\perp)\equiv\int d^2{\imb x}_\perp\, e^{i{\imb x}_\perp\cdot{\imb k}_\perp}
e^{-B_2({\imb x}_\perp)}=\int d^2{\imb x}_\perp e^{i{\imb x}_\perp\cdot{\imb k}_\perp} \left<U(0)U^\dagger({\imb x}_\perp)\right>\; .
\end{equation}
This object depends only on the modulus of ${\imb k}_\perp$ thanks to
the rotation invariance of the function $B_2$.

It is rather easy to obtain the first two terms of the asymptotic
expansion of $C(k_\perp)$ for $k_\perp\gg Q_s$. Indeed, for such a
large $k_\perp$, the typical $x_\perp$ contributing in the integral is of
order $1/k_\perp$, and $Q_s x_\perp\ll 1$. We are therefore in the region
where the function $B_2({\imb x}_\perp)$ is know analytically through
formula (\ref{eq:B2-approx}). The asymptotic expansion of $C(k_\perp)$
is then obtained by expanding the exponential of $-B_2$, which gives
\begin{equation}
C({\imb k}_\perp)=
2{{Q_s^2}\over{k_\perp^4}}+{8\over\pi}{{Q_s^4}\over{k_\perp^6}}
\left(\ln\left({{k_\perp}\over{\Lambda_{_{QCD}}}}\right)-1\right)+{\cal O}\left({{Q_s^6}\over{k_\perp^8}}\right)\; .
\label{eq:Ck-asympt}
\end{equation}
One can also notice that the correlator $C(k_\perp)$
obeys the sum rule
\begin{equation}
\int {{d^2{\imb k}_\perp}\over{(2\pi)^2}} C(k_\perp)=e^{-B_2(0)}=1\; .
\end{equation}

\begin{figure}[ht]
\centerline{
\resizebox*{!}{7.5cm}{\rotatebox{0}{\includegraphics{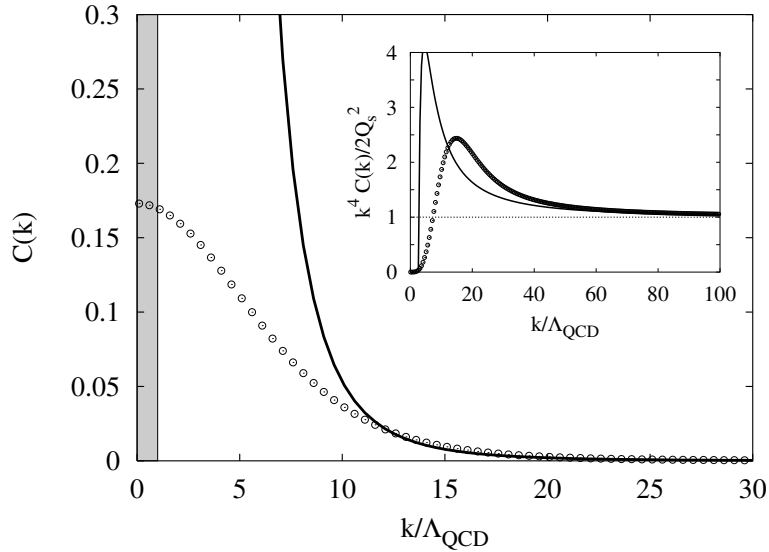}}}}
\caption{\label{fig:Ck}Value of $C(k_\perp)$ as a function of
$k_\perp/\Lambda_{_{QCD}}$ for $Q_s/\Lambda_{_{QCD}}=10$. The open
circles are the result of a numerical evaluation, and the solid line
is the asymptotic formula given in Eq.~(\ref{eq:Ck-asympt}). The
shaded area indicates the region where $k_\perp\le \Lambda_{_{QCD}}$.
One can see on the numerical curve the onset of saturation at low
values of $k_\perp$. In the upper right corner, we have plotted the
same data for the quantity $k_\perp^4 C(k_\perp)/2Q_s^2$, which allows
for an easier reading for large values of $k_\perp$.}
\end{figure}

Numerical results as well as the asymptotic formula for $C(k_\perp)$
are displayed in figure \ref{fig:Ck}.

\eject

\end{document}